\documentclass[aps,prc,twocolumn,showpacs,superscriptaddress]{revtex4-2}
\usepackage{natbib}
\usepackage{graphicx}
\usepackage{color}
\usepackage{float}
\usepackage{natbib}
\usepackage{graphicx}
\usepackage[usenames,dvipsnames]{xcolor}
\usepackage{amsfonts}
\usepackage{amsmath,amssymb}
\usepackage{multirow,bigdelim}
\usepackage{dcolumn}
\usepackage{bm}
\usepackage{pifont}
\usepackage{wasysym}
\usepackage{oplotsymbl}
\usepackage{xcolor}
\usepackage{float}
\usepackage{hyperref}
\definecolor{MyDarkTeal}{HTML}{007171}
\definecolor{purple1}{HTML}{830a88}
\definecolor{23}{HTML}{ff00ff}
\definecolor{24}{HTML}{5e81b5}
\definecolor{25}{HTML}{8c0000}

\begin{document}

\title{Systematics of characteristics of pygmy dipole resonances in medium-heavy and heavy atomic nuclei with neutron excess}

\author{V. A. Plujko}
\email[]{plujko@gmail.com}
\affiliation{Taras Shevchenko National University of Kyiv, 01601 Kyiv, Ukraine}

\author{O.M. Gorbachenko}
\affiliation{Taras Shevchenko National University of Kyiv, 01601 Kyiv, Ukraine}

\author{N.O. Romanovskyi}
\affiliation{Taras Shevchenko National University of Kyiv, 01601 Kyiv, Ukraine}

\date{\today}

\begin{abstract}

The systematics of energies and the contribution of pygmy dipole resonance (PDR) to the energy-weighted sum rule of dipole gamma transitions
in medium-heavy and heavy nuclei with an excess of neutrons are considered. The modified macroscopic model of Isacker-Nagarajan-Warner
was used for calculating PDR energies with the number of surface neutrons proportional to the thickness of the neutron skin according to
the Pethick- Ravenhall expression (PR INW approach). 
Such modification of the  macroscopic approach  by Isacker-Nagarajan-Warner enables to take into account microscopic evidence  of direct relationship between skin thickness and low-energy  dipole response.
The results are compared with the microscopic calculations for the chains of
Ni, Sn and Pb isotopes. 
It was demonstrated that the dependence of the magnitudes of the energies within the PR INW approach on neutron excess is in rather good agreement with experimental data and microscopic calculations 
if the absolute value of the strength of the neutron-proton interaction is nearly three times as large   as that obtained by Isacker-Nagarajan-Warner by the volume integral of the nucleon-nucleon interaction. 
While the macroscopic INW PR model can describe the main features of the PDR, above mentioned discrepancy of the strength values doesn't not provide reason enough for the conclusion that PDR is pure collective state.

The analytical expressions for the PDR fraction of the energy-weighted sum rule for electric dipole transitions (E1 EWSR) are used.
They are based on the "molecular" energy-weighted E1 sum rule considering the number of surface neutrons as a function of the neutron
thickness (PR MSR approach). Systematics for the PDR fraction of E1 EWSR are proposed with parameters obtained by fitting the experimental
data and microscopic calculations.

\end{abstract}

\pacs{ 
      21.60.Ev,    
      21.60.-n,     
      24.30.Cz ,    
      24.30.Gd     
           }

\maketitle

\section{Introduction}
\label{introduction}

In recent years, much attention has been paid to experimental and theoretical studies of collective states of atomic nuclei, in particular, the pygmy dipole resonance, at energies near the neutron separation energy \cite{Savran2013,Savran2015,Zilges2015,Bracco2019,Lanza2023a,Lanza2023b,Lanza2023c}.

Although PDR fraction in the energy-weighted  sum rule usually does not exceed few percents, it can have significant influence on various nuclear processes
\cite{Brzosko1969,Igashira1986,Goriely1998,Arnould2007,Goriely2023,Aumann2019,Zilges2022}, in particular on the rate of neutron absorption in
the r-process which is important for analyzing the distribution of elements in the Universe. The account of the PDR  significantly
improves the agreement of theoretical calculations and experimental values of the distribution of elements in the Universe \cite{Goriely1998,Arnould2007,Goriely2023}.

There are various microscopic and macroscopic methods for calculation of the properties of the strength functions with PDR excitation and PDR characteristics
\cite{Lanza2023a,Goriely2023,Chambers1994,Vretenar2001,Goriely2002,Paar2005,Paar2003,Piekarewicz2006,Tsoneva2008,Litvinova2009,Co2013,Lanza2015,Abrosimov2009,Suzuki1990,Isacker1992,Baran2012a,Baran2012b,Croitoru2015,Plujko2017,Gorbachenko2023}.
Specifically it was demonstrated that the PDR in the neutron excess nuclei is a mode directly related to the size of the neutron skin \cite{Paar2005,Tsoneva2008}.

For the first time the analytical expression for the PDR energy was obtained by Suzuki-Ikeda-Sato \cite{Suzuki1990} in line with the hydrodynamic model by the Steinwedel-Jensen for the description of the giant dipole resonances (GDR). The PDR was considered as an out-of-phase density vibration in the nuclear volume of two liquids that corresponds to nucleons in the core of the nucleus and the excess neutrons in the skin (SIS model).  The expression for the PDR energy based on the Goldhaber-Teller model\cite{Goldhaber1948} was obtained by Isacker-Nagarajan-Warner in  Ref.\cite{Isacker1992} (INW model).  In this approach, PDR mode corresponds to out-of-phase displacement of the surface density against the neutron--proton core.  Nucleon densities were taken  in the  form of the Fermi  function with two parameters (2pF distribution).

The dynamics of the dipole oscillations are governed by the neutron--proton interaction. Here we will use the expression by Pethick- Ravenhall (PR approach)\cite{Pethick1996} for the calculation of the number of surface neutrons as a linear function of the thickness of the neutron skin. This approach is called below as the PR  INW model.
This modification enables to take into account microscopic observation of direct relationship between skin thickness and low-energy  dipole response, and approach is called below as the PR INW model.

The thickness of the neutron skin is approximated by the linear function of the  neutron-proton asymmetry   parameter $I=(N-Z)/A$ with coefficients that  depend on the mass numbers of  the atomic nucleus. The energies of PDR for isotopes Ni, Sn, and Pb are calculated within the PR INW model.  It is shown that, in general, the results of the calculations are consistent with  the microscopic calculations after fitting the constants of  the neutron--proton interaction.

The PDR energies were also calculated within the SIS model using the expression by Pethick- Ravenhall (PR SIS approach) \cite{Pethick1996} for the number $N_{s}$ of surface neutrons.  General behavior of PDR energies by the PR SIS model as a function of neutron-proton asymmetry contradicts to the results of the PR INW model as well as the microscopic calculations. Such behavior of $E_{p}$ for the SIS-model was already demonstrated in Refs.\cite{Chambers1994,Vretenar2001} at $N_{s} =N-Z$.

In the  present contribution the analytical expressions for the PDR fraction of the E1 EWSR are used. They are based on the "molecular" energy-weighted E1 sum rule (MSR) with the number of surface neutrons considered as function of the neutron thickness (PR MSR approach). Systematics for the PDR fraction of E1 EWSR are proposed with the parameters obtained by fitting experimental data and microscopic calculations.

The paper is organized as follows.  In Sec. II the analytical  expression for PDR energy as a  function of neutron skin is derived within the framework of the PR INW approach. The simple analytical expressions for neutron skin are presented.  The results of the PDR energy calculations within the framework of the macroscopic PR INW and PR SIS  models are compared with both  the microscopic calculations and experimental data. Systematics of the values of the PDR energies are given by fitting the microscopic and experimental results.

The PDR fraction of the dipole energy-weighted sum rule   is considered within the framework of the macroscopic expression in Sec. III. Systematics of this fraction is presented and compared with the experimental data.  Section IV contains the summary and  the conclusions.  In the Appendix the procedures for the calculations of the PDR energy within the  PR INW and PR SIS models as well as nuclear geometric quantities are outlined.

\section{Comparison of PDR energies and systematics}
\label{ComparisonPDR}

\begin{figure}[htb]
\centering
\setlength{\abovecaptionskip}{0pt}
\setlength{\belowcaptionskip}{8pt}
\includegraphics[scale=0.6]{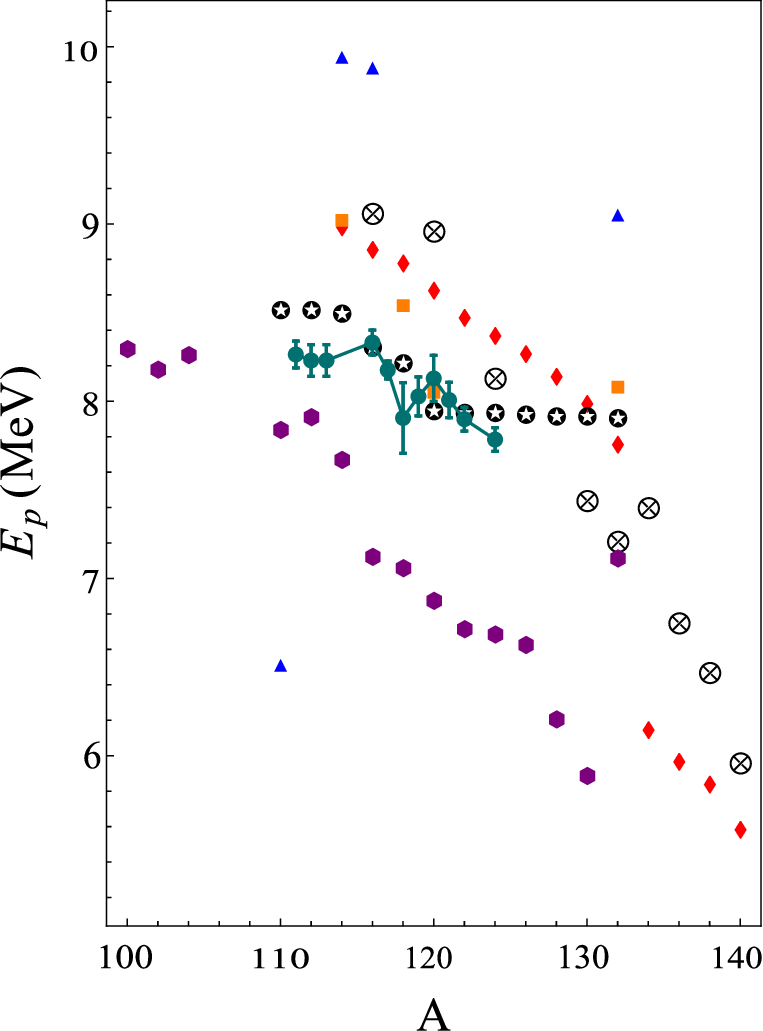}
\caption{
The energies of the PDR for isotopes of tin ${}_{50}^{A} Sn$  with $N\ge Z$ in dependence on the atomic mass number.  The symbols correspond to the microscopic calculations by the following methods: square $\color{orange}\blacksquare$ -  RRPA\cite{Vretenar2001}, rhombus $\color{red}\blacklozenge$ - RQRPA\cite{Paar2005}, circle with an asterisk \ding{74} - MF RRPA\cite{Piekarewicz2006}, hexagon $\color{violet} \hexagofill$ -  QPM \cite{Tsoneva2008}, circle with a cross $\otimes$ - RQTBA\cite{Litvinova2009}, triangle $\color{blue}\blacktriangle$ - HF CRPA\cite{Co2013}. The experimental data are indicated by the shaded circle $\color{MyDarkTeal}\CIRCLE$. They are connected by the lines for the better visualization of the data. These data are taken from  Fig. 2 of Ref.\cite{Markova2025} and correspond to the average values of the PDR energies.
}
\label{fig:fig1}
\end{figure}

 Calculated values of the PDR energy are somewhat different in the framework of different microscopic methods.  For example, the values of the energy $E_{p}$ according to microscopic calculations presented in Refs.\cite{Vretenar2001,Paar2005,Piekarewicz2006,Tsoneva2008,Litvinova2009,Co2013}  for isotopes of tin are shown in figure  \ref{fig:fig1}. The following microscopic methods were used: the relativistic RPA (RRPA)\cite{Vretenar2001}, the relativistic quasiparticle RPA (RQRPA) \cite{Paar2005}, mean-field (MF) plus the relativistic RPA (MF RRPA)\cite{Piekarewicz2006}, the quasiparticle-phonon model (QPM)\cite{Tsoneva2008}, the relativistic quasiparticle time blocking approximation (RQTBA)\cite{Litvinova2009}, a self-consistent Hartree-Fock plus continuum RPA approach (HF CRPA)\cite{Co2013}.

The figure \ref{fig:fig1} shows that microscopic values of the PDR energies decrease with growing parameter of the neutron-proton asymmetry.  Calculated values of the PDR energies agree within $\approx$ 30\%.
It should be noted  that the microscopic value of the PDR energies is, in fact, the energy of the peak of the strength function in the range of the low-energy peak. The magnitude of this energy depends on many calculation parameters, for example, on the number and character of the single-particle states that are involved in the calculations as well as on coupling forces between PDR and other states. Uncertainties in microscopic values of the energy in Fig.1 may be conditioned by these peculiarities of the microscopic calculations.  We compare these results of the microscopic approaches  with the results of macroscopic calculations within the framework of PR INW and PR SIS approaches. In macroscopic approaches  the quantity  $E_{p}$   is the energy of the PDR state, and  fragmentation of the resonance state is usually considered as a broadening due to the width  resulting from one body dissipation.

The expression for the PDR energy in the INW approach\cite{Isacker1992} is given by Eq.(\ref{eq:A2}) in the Appendix. As mentioned before, this method includes two parameters: 1)  the number of surface neutrons $N_{s} $, and 2) the thickness $y$ of the neutron skin for 2pF Fermi nuclear distributions (\ref{eq:A4}), i.e., a difference between neutron and proton radii of the distributions. We use expression by Pethick- Ravenhall (\ref{eq:A17}): $N_{s} =3yN/R_{0} $, for dependence  thickness of surface neutron number  on the neutron skin and Eq.(\ref{eq:A10}) for calculation  of quantity $y$  as a linear function of the root-mean-square thickness $\Delta r_{np}$: $y=\sqrt{5/3} \cdot \Delta r_{np} $. Therefore, we have the following relationship for PDR energy within the PR INW  approach

\begin{multline}
{E_p}= K_p\left[\dfrac{1}{3} \dfrac{Z}{A-N_s}\right]^{1 / 2}\left[\dfrac{6 a}{y R_m} \Delta F\left(R_n, R_p\right)\right]^{1 / 2} \times \\ \times A^{-1 / 6}
 \equiv
E_{p,1}
\label{eq:1}
\end{multline}

\noindent where $\Delta F=F(R_{n} ,R_{p} )-F(R_{p} ,R_{p} )$ with functions $F(R_{p} ,R_{p} )$, $F(R_{n} ,R_{p} )$ that describe the displacement between 2pF distributions of neutrons and protons. The quantities$R_{p} =R_{m} -y/2$, $R_{n} =R_{m} +y/2$ are the radii of the 2pF distributions for protons and neutrons,  $R_{m} =(R_{p} +R_{n} )/2$  is average radius,  $R_{0} =r_{0} A^{1/3} $ is the radius of nuclear distributions with a sharp edge. The factor $K_{p} $ is given by the equation

\begin{equation}
K_{p} =\left[\frac{\hbar ^{2} }{m} \frac{|\kappa _{np} |}{8\pi ar_{0}^{4} } \frac{r_{m} }{r_{0} } \right]^{1/2} =1.34\cdot \sqrt{|\kappa _{np} |} (MeV),
\label{eq:2}
\end{equation}

\noindent where $|\kappa _{np} |$  is the absolute value of the strength of effective neutron- proton interaction;  $m$ - nucleon mass, $a$ - the diffuseness parameter of 2pF distributions which is taken identical for protons and neutrons, $r_{m} =R_{m} /A^{1/3} $.

 In accordance with Eq.(\ref{eq:A9}), the expression  for  PDR energy has the following form  at  $R_{m} >>\pi a$

\begin{multline}
E_{p} =K_{_{p} } \left[\frac{1}{3} \frac{Z}{A-N_{s} } \right]^{1/2} \left(1-\frac{y\cdot R_{m} }{10 a^{2} } \right)^{1/2}
\times \\ \times
A^{-1/6} \equiv E_{p,2}.
\label{eq:3}
\end{multline}

\noindent and in a linear approximation in $y$ of the function $\Delta F(R_{n} ,R_{p} )/y$ . Therefore we have for small values of $y/a$  the following equation

\begin{multline}
E_{p} \cong K_{p} \left[\frac{1}{3} \frac{Z}{A-N_{s} } \right]^{1/2} \left(1-\frac{y\cdot R_{m} }{20a^{2} } \right)
\times \\ \times
A^{-1/6} \equiv E_{p,3}.
\label{eq:4}
\end{multline}

The dependence of PDR energy on the quantities in the brackets given by Eq.(\ref{eq:4}) is the same as for the INW model ( Eq.(\ref{eq:10}) of Ref.\cite{Isacker1992} in the leptodermic approximation $R_{0} >>\pi a$ ).

\begin{figure}[htb]
\centering
\setlength{\abovecaptionskip}{0pt}
\setlength{\belowcaptionskip}{8pt}
\includegraphics[scale=0.65]{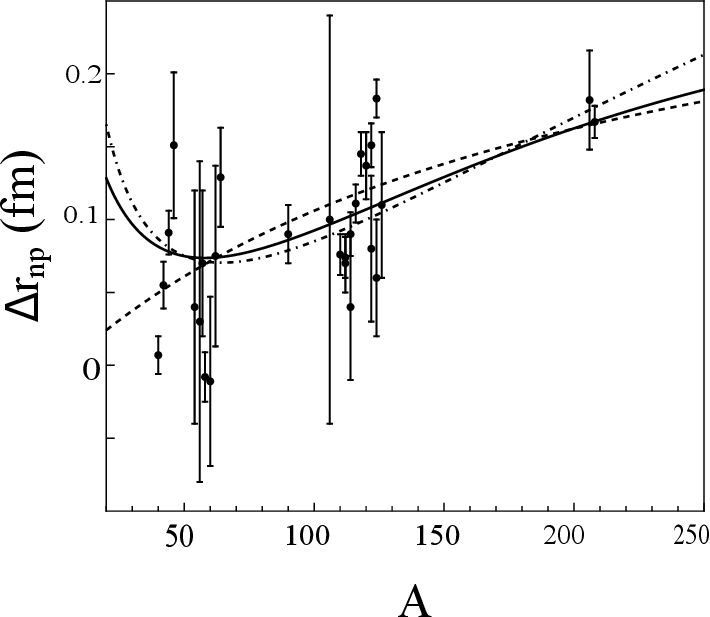}
\caption{
The dependence of the rms thickness of  the neutron skin $\Delta r_{np} $ on the atomic mass number along the beta-stability line. Denotations of the curves: dashed line - for parameters(\ref{eq:A16a});  dot-dashed line - for parameters (\ref{eq:A16b}); and  solid line - for parameters (\ref{eq:A16c}).  The experimental values are taken from  Refs.\cite{Trzcinska2001,Jastrzebski2004,Zhang2021} for isotopes near beta-stability line.}
\label{fig:fig2}
\end{figure}

 Skin thickness $y$ for 2pF distributions  is determined by Eq.(\ref{eq:A13}) as a linear function of the parameter  neutron-proton asymmetry $I=(N-Z)/A$: $\Delta r_{np} =\alpha +\beta \cdot I$. The parameters $\alpha ,\, \, \beta $ are given by the Eqs.(\ref{eq:A14})-(\ref{eq:A16c}).  The results of the calculations of rms neutron skin with these parameters along the beta-stability line are shown in Fig.\ref{fig:fig2}.  The beta-stability line is defined by Green formula \cite{Green1951,Green1952}: $I=0.4A/(200+A)$.

It can be seen from the figure that the values of the rms thicknesses with different parameters are in rather close agreement and they are not contradictory to experimental data. In what follows the expression for $\Delta r_{np} $ with parameters (\ref{eq:A15}), (\ref{eq:A16c}) is adopted, because it fits the data with the minimal chi-square deviation, i.e., the following expression is taken
\[\Delta r_{np} =0.62\, {\rm -}\, {\rm 0.22}A^{1/3} +{\rm 0.003\; }A+(0.47A^{1/3} {\rm -}\, {\rm 0.009}A)\cdot I .\]

The factor $K_{p} $, Eq.(\ref{eq:2}), is calculated using the following values of the parameters: $a=0.57$~fm, $r_{0} $=1.15~fm \cite{Myers1983}, $r_{m} =1.25$~fm (\ref{eq:A21}), $\hbar ^{2} /m\simeq \hbar ^{2} /m_{proton} $$\simeq 41.5$${\rm MeV}\cdot {\rm fm}^{{\rm 2}} $ . For the value of the effective proton-neutron interaction, we take the value of the constant component $t_{0} $ of the central interaction of the Skyrme force  MSk7\cite{Goriely2001}, $\kappa _{np} = t_{0} =-1828.23\, \, {\rm MeV}\cdot {\rm fm}^{{\rm 3}}$ with the result

\begin{equation}
K_{p} =57.3 ~{\rm MeV.} \label{eq:5}
\end{equation}

\begin{figure}[htb]
\centering
\setlength{\abovecaptionskip}{0pt}
\setlength{\belowcaptionskip}{8pt}
  \includegraphics[scale=0.48]{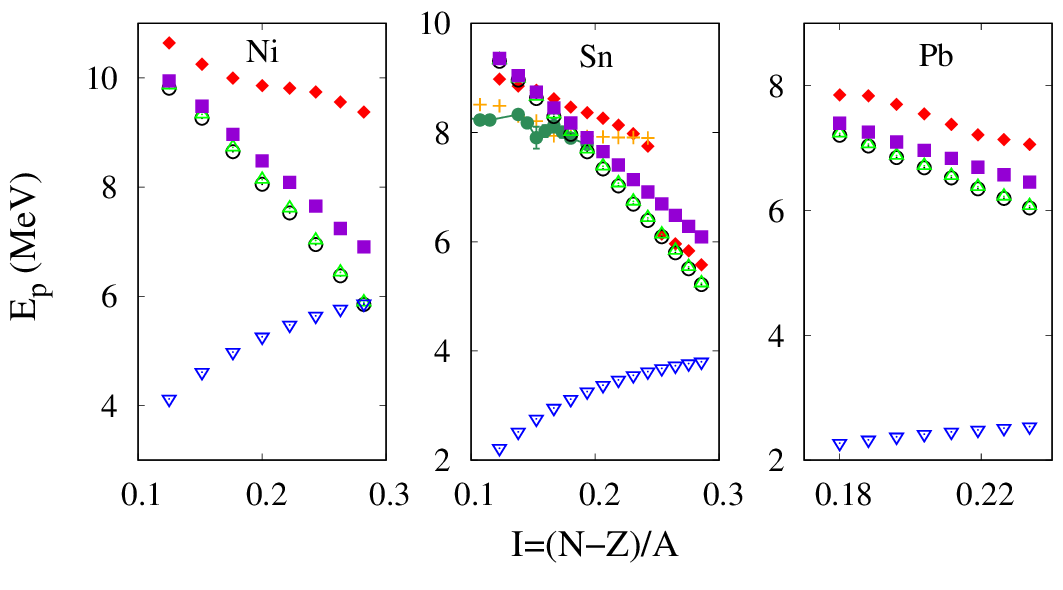}
\caption{
The comparison of PDR energy values for  different theoretical methods and experimental data as a function of the parameter of neutron-proton asymmetry $I=(N-Z)/A$ for isotopes Ni, Sn, Pb. Microscopic models: 
rhombus $\color{red}\blacklozenge$ - RQRPA\cite{Paar2005}, 
cross $\color{orange}+$ - MF RRPA\cite{Piekarewicz2006};  
experimental data - filled circle $\color{MyDarkTeal}\CIRCLE$ that correspond to the average PDR energies from Fig. 2 of \cite{Markova2025}. 
Macroscopic approaches:  
circle $\odot$ - PR INW ($E_{p,1} $, Eq.(\ref{eq:1})),  
triangle $\color{green} \triangle$ - approximation (\ref{eq:3}) for PR INW ($E_{p,2} $); 
square $\color{purple1} \blacksquare$ - approximation (\ref{eq:4}) for PR INW  ($E_{p,3} $), 
inverted triangle $\color{blue}\nabla$ - PR SIS  Eq.(\ref{eq:A26}) with PR approach for $N_{s}$.
}
\label{fig:fig3}
\end{figure}

Figure \ref{fig:fig3} demonstrates comparison between the PDR energies calculated for isotopes of  Ni, Sn, Pb within microscopic relativistic quasiparticle RPA (RQRPA)\cite{Paar2005} and the results of the macroscopic PR INW and PR SIS models. For the PR INW model,  calculations are performed by Eqs.(\ref{eq:1})-(\ref{eq:3}),(\ref{eq:5}). Energies of the SIS model are calculated using Eq.(\ref{eq:A26}) with the PR expression for the number of surface neutrons  $N_{s} $ (PR SIS model). Experimental data from Ref.\cite{Markova2025} for isotopes Sn are also shown as well as the results of microscopic calculations within the  mean-field  plus the relativistic RPA (MF RRPA) \cite{Piekarewicz2006}.

 Figure  \ref{fig:fig3} shows that  the values of the PDR energies calculated within the macroscopic approach PR INW are rather close to the calculations of the energies by microscopic relativistic RRPA and MF RPA in atomic nuclei with mass numbers A$>\approx$100. Deviation of the macroscopic energies within PR INW from microscopic values does not exceed uncertainties of calculations within different microscopic methods ($\approx$30\%).
 The results within the framework of these  approaches for isotopes of Sn as a whole are not too different from the experimental data.  Dependence of the PDR energies on  neutron-proton asymmetry  for the energies within the PR SIS model is contradictory to the results of microscopic calculations, as it was early demonstrated for  $N_{c} =Z$ in Refs.\cite{Chambers1994,Vretenar2001}.

For the estimation of the microscopic values of  PDR energies for atomic nuclei in the absence of calculations, systematics can be used. For this systematics, we will use the following expressions

\begin{equation}
E_{p} =e_{1} \sqrt{Z} A^{-2/3} /\sqrt{1+e_{2} \Delta r_{np} A^{1/3} } =E_{p,4},
\label{eq:6}
\end{equation}

\begin{multline}
E_{p} =d_{1} \sqrt{Z} A^{-2/3} /\sqrt{1+d_{2} \Delta r_{np} A^{1/3} } +d_{3} \left[\Delta r_{np} \right]^{1/2}
 \times \\ \times
 A^{-5/6} =E_{p,5}.
 \label{eq:7}
\end{multline}

The first expression is based on the PR INW model, the second one is based on the sum of PR INW and PR SIS approaches. The parameters $e_{j} $, $d_{j} $  are obtained to fit values of microscopic data for the PDR energies to Eqs.(\ref{eq:6}), (\ref{eq:7}). The least squares method  is used for the fitting with minimization of  the chi-square deviation of the form $\chi _{_{\alpha } }^{2} $ = $\sum _{j=1}^{N_{\Sigma } }w_{j}  \left(E_{p,theor} (j)-E_{p,\alpha } (j)\right)^{2} $, $\alpha =1-5$. The $N_{\Sigma } $ is the total amount of the data-points. The weight factors $w_{j} $ are the contributions of the square deviations of  the fitted PDR energy  $j$ from its microscopic value. Microscopic data were taken from calculations within  the following methods. RRPA\cite{Vretenar2001},  RQRPA\cite{Paar2005}, HB RQRPA\cite{Paar2003}, MF RRPA \cite{Piekarewicz2006}, QPM \cite{Tsoneva2008}, RQTBA \cite{Litvinova2009}, HF CRPA \cite{Co2013}. We use the constant weight factors $w_{j} =1\, \, {\rm MeV}^{{\rm -2}} $, because  systematic uncertainties of the microscopic calculations are not easily specified.

The values of the fitted parameters are given in Table \ref{tab:tab1} together with the ratio of the chi-square value for the given method $\chi _{\alpha }^{2} $ to the $\chi _{1}^{2} $ for PR INW model  with  $\kappa _{np} =- 1828.23\, \, {\rm MeV}\, \cdot {\rm fm}^{{\rm 3}}$: $\bar{\chi }_{1}^{2} \equiv \chi _{1}^{2} /N_{\Sigma } $ = 2.38.

\begin{table}[ht]
\centering
\caption{
The parameters of systematics of microscopic calculations  of the PDR energies  $E_{p,\alpha } $, $\alpha =1-5$, and ratio of the value of the chi-square deviation for given approach  to that one for PR INW model  at $\kappa _{np} =-\, 1828.23\, $ ${\rm MeV}\, \cdot {\rm fm}^{{\rm 3}} $; $\bar{\chi }_{1}^{2} \equiv \chi _{1}^{2} /N_{\Sigma } $ = 2.38.
}
\label{tab:tab1}
\begin{ruledtabular}
\begin{tabular}{ccc}
Expression & Parameters  & $\bar{\chi }_{E_{p,\alpha } }^{2} /\bar{\chi }_{E_{p,\alpha =1} }^{2} $ \\ \hline
$E_{p,1}$, (\ref{eq:1}) & $\kappa_{np} =-1828.23~ {\rm MeV} \cdot {\rm fm}^{{\rm 3}} \, $ \cite{Goriely2001} & 1 \\
 & $\kappa _{np} =-1930\pm  78 ~{\rm MeV} \cdot {\rm fm}^{{\rm 3}} \, $ & 0.98 \\
$E_{p,2} $, (\ref{eq:3}) & $\kappa _{np} =-1828.23~{\rm MeV} \cdot {\rm fm}^{{\rm 3}} $ \cite{Goriely2001} & 0.98 \\
$E_{p,3} $, (\ref{eq:4}) & $\kappa _{np} =-1828.23~{\rm MeV}\cdot {\rm fm}^{{\rm 3}} \, \, $\cite{Goriely2001} & 0.66 \\
$E_{p,4} $, (\ref{eq:6}) & $e_{1} ={\rm 27.3}$, $e_{2} ={\rm -0.03}$ & 0.38 \\
$E_{p,5} $, (\ref{eq:7}) & $d_{1} ={\rm 27.1}$, $d_{2} ={\rm 0.01}$, $d_{3} ={\rm 31.1}$ & 0.37 \\
\end{tabular}
\end{ruledtabular}
\end{table}

\begin{figure}[htb]
\centering
\setlength{\abovecaptionskip}{0pt}
\setlength{\belowcaptionskip}{8pt}
\includegraphics[scale=0.48]{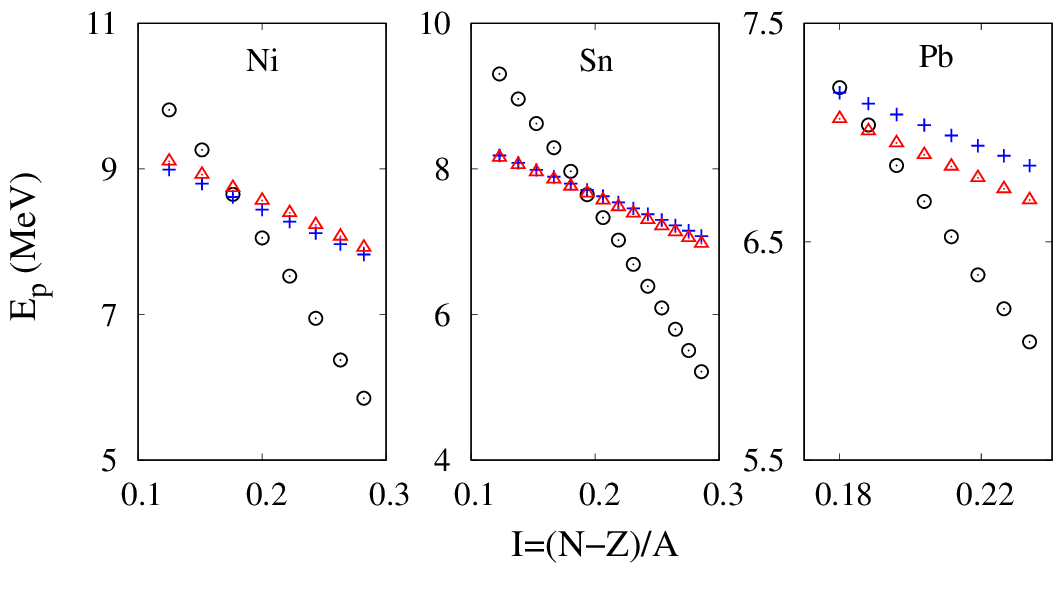}
\caption{
Comparison of the systematics of the PDR energy with the results of the PR INW model for isotopes Ni, Sn, Pb in dependence on the parameter of the neutron-proton asymmetry.  
Denotations: circle $\odot$ - PR INW model ($E_{p,1} $), 
cross $\color{blue}+$ - microscopic data systematics $E_{p,4} $, Eq.(\ref{eq:6}), 
triangle $\color{red} \triangle$ - microscopic data systematics $E_{p,5} $, Eq.(\ref{eq:7}).
}
\label{fig:fig4}
\end{figure}

It is seen that Eqs. (\ref{eq:6}), (\ref{eq:7}) can be considered as the best systematics of microscopic calculations of the PDR energy in medium-heavy and heavy nuclei with neutron-excess.  One can recognize that the value of the proton-neutron interaction $\kappa _{np} =-1930\pm  78\, {\rm MeV}\, \cdot {\rm fm}^{{\rm 3}} $ from fitting the PR INW model to microscopic data agrees  rather closely with the value of the constant component $t_{0} $$=-1828.23\, \, {\rm MeV}\cdot {\rm fm}^{{\rm 3}} $ of the central interaction of the Skyrme force  MSk7 \cite{Goriely2001},

Figure \ref{fig:fig4} demonstrates comparison of the microscopic data systematics  of the PDR energy given by Eqs. (\ref{eq:6}), (\ref{eq:7}) with the results of the PR INW model for isotopes Ni, Sn, Pb in dependence on the parameter of the neutron-proton asymmetry $I=(N-Z)/A$.

One can see that the values of the PDR energies within microscopic data systematics (\ref{eq:6}), (\ref{eq:7}) are in very close agreement.   The PDR energies by systematics fall more slowly with the growth of the neutron-proton asymmetry than that within the PR INW model.

\section{PDR fraction of the E1 EWSR}

Another important characteristic of PDR excitation  is the fraction of PDR ($\sigma _{{\rm int}} {\rm (PDR)}$) of the integrated photo-absorption cross-section  $\sigma _{int} $ for electric dipole gamma quanta. This component of cross-section  $\sigma _{{\rm int}} {\rm (PDR)}$ is proportional to the PDR fraction ($m_{{\rm L=1}}^{1} {\rm (PDR)}$) of the E1 EWSR $m_{{\rm L=1}}^{{\rm 1}} $ \cite{Lipparini1989}. The ratio of  the PDR to GDR sum rule is proportional to the number of  neutrons $N_{s} $ in the neutron skin (\cite{Suzuki1990,Isacker1992,Alhassid1982,Kurasawa1995,Lipparini1989})

\begin{multline}
\bar{s}_{p} =\frac{m_{{\rm L=1}}^{1} {\rm (PDR)}}{m_{{\rm L=1}}^{1} {\rm (GDR)}} =\frac{\sigma _{{\rm int}} {\rm (PDR)}}{\sigma _{{\rm int}} {\rm (GDR)}} =\frac{Z}{N} \frac{N_{s} }{A-N_{s} }  .
\label{eq:8}
\end{multline}

This relationship corresponds to the ratio with "molecular" energy-weighted E1 sum rule for the PDR fraction \cite{Alhassid1982,Kurasawa1995,Lipparini1989}. It is denoted below as the MSR expression.

With allowance for Eq. (\ref{eq:A17}) for a number near surface neutrons, this equation equals to (expression PR MSR)

\begin{multline}
\bar{s}_{p} =\frac{\sigma _{{\rm int}} {\rm (PDR)}}{\sigma _{{\rm int}} {\rm (GDR)}} =\frac{m_{{\rm L=1}}^{1} {\rm (PDR)}}{m_{{\rm L=1}}^{1} {\rm (GDR)}} = 
\\
=\frac{Z}{A-N\cdot \sqrt{15} \cdot \Delta r_{np} /R_{0} } \frac{\sqrt{15} \cdot \Delta r_{np} }{R_{0} } =\bar{s}_{p,1}.
 \label{eq:9}
 \end{multline}

In the leptodermic approximation and in the case $N\cong Z\cong A/2$ formula (\ref{eq:9}) takes the form
\begin{multline}
\bar{s}_{p,1} \cong \frac{1}{2-\cdot \sqrt{15} \cdot \Delta r_{np} /R_{0} }  \frac{\sqrt{15} \cdot\Delta r_{np} }{R_{0} } \approx
\\
 \approx1.68\cdot \frac{\Delta r_{np} }{A^{1/3} } =\bar{s}_{p,2}.
 \label{eq:10}
\end{multline}

We suppose, that  E1 EWSR (integral cross-section $\sigma _{int} $) can be taken to be the sum of two components  from  excitations of  the GDR and PDR. Therefore, the expression for the PDR fraction of the E1 EWSR within PR MSR is considered in the form

\begin{multline}
s_{p} =\frac{m_{{\rm L=1}}^{1} {\rm (PDR)}}{m_{{\rm L=1}}^{1} } =\frac{\sigma _{{\rm int}} {\rm (PDR)}}{\sigma _{{\rm int}} } =
\\
=\frac{\bar{s}_{p,1} }{1+\bar{s}_{p,1} } =s_{p,1}.
 \label{eq:11}
\end{multline}

 For estimation of the fractions of PDR to GDR in E1 EWSR (integral cross-section $\sigma _{int} $) for atomic nuclei in the absence of theoretical calculations or experimental data, systematics can be used. For systematics, we will use the following expressions

\begin{multline}
\bar{s}_{p} =\frac{\sigma _{{\rm int}} {\rm (PDR)}}{\sigma _{{\rm int}} {\rm (GDR)}} =\frac{m_{{\rm L=1}}^{1} {\rm (PDR)}}{m_{{\rm L=1}}^{1} {\rm (GDR)}} =
\\
=\delta \left(\frac{\Delta r_{np} }{A^{1/3} } \right)^{\gamma } \equiv \bar{s}_{p,sys}.
\label{eq:12}
\end{multline}

Equation (\ref{eq:12}) is similar to Eq.(\ref{eq:10}) regarding dependence on $\Delta r_{np} /A^{1/3} $. By analogy with Eq.(\ref{eq:11}),  we take the following relationship for the systematics of the PDR fraction of E1 EWSR (integral cross-section $\sigma _{int} $)

\begin{multline}
s_{p,sys} =\frac{m_{{\rm L=1}}^{1} {\rm (PDR)}}{m_{{\rm L=1}}^{1} } =\frac{\sigma _{{\rm int}} {\rm (PDR)}}{\sigma _{{\rm int}} }=
\\
 =\frac{\bar{s}_{p,sys} }{1+\bar{s}_{p,sys} }.
\label{eq:13}
\end{multline}

 The parameters $\delta $, $\gamma $ in Eq.(\ref{eq:12}) are obtained from fitting the values of microscopic or experimental data for ratio $\bar{s}_{p} $ by systematics $\bar{s}_{p,sys} $. The least squares method  is used for the fitting with minimization of  the chi-square deviation $\chi _{\bar{s}_{p,sys} }^{2} $ = $\sum _{j=1}^{\bar{N}_{\Sigma } }w_{j}  \left(\bar{s}_{p,data} (j)-\bar{s}_{p,sys} (j)\right)^{2} $. Constant weight factors $w_{j} =1/10^{-4} $ are taken. The values of the parameters $\delta $, $\gamma $ are found from the fitting results of microscopic calculations and experimental data separately. The theoretical data of the following microscopic methods are used - RRPA \cite{Vretenar2001}, HB RQRPA \cite{Paar2003}  and MF RRPA\cite{Piekarewicz2006}.
 Experimental data are taken from Fig.26 of Ref.\cite{Savran2013} and Fig.2 of Ref.\cite{Markova2025}.  Data in Ref.\cite{Savran2013} were collected from  Refs.\cite{Govaert1998,Adrich2005,Savran2011,Hartmann2004,Klimkiewicz2007,Schwengner2007,Schwengner2008,Ozel2007,Volz2006,Makinaga2010,Schwengner2010,Schramm2012,Wieland2009,Poltoratska2012,Iwamoto2012,Toft2011}. All these experimental data correspond to the fraction $s_{p,data} $ of PDR to E1 EWSR. Before fitting, we recalculate $s_{p,data} $ to $\bar{s}_{p,data} $ by the relationship $\bar{s}_{p,data} =s_{p,data} /(1-s_{p,data} )$. Fitted parameters are presented in Table 2.  The last column in Table 2 shows the ratio of the chi-square values $\chi _{\bar{s}_{p,k} }^{2} $ for ratios $\bar{s}_{p,sys} $, $\bar{s}_{p,2} $ to the value $\chi _{\bar{s}_{p,1} }^{2} $  ($\bar{\chi }_{\bar{s}_{p,1} }^{2} =\chi _{\bar{s}_{p,1} }^{2} /\bar{N}_{\Sigma } $ =2.8 (4.0)) for  analytical expression PR MSR (\ref{eq:9}).

\begin{table}[ht]
\centering
\caption{
Parameters of systematics (\ref{eq:12}) for the ratio of fractions of PDR to GDR in E1 EWSR from fitting the results of the microscopic calculations within RRPA\cite{Vretenar2001}, HB RQRPA \cite{Paar2003}, MF RRPA\cite{Piekarewicz2006} as well as  experimental data from Refs.\cite{Savran2013}, \cite{Markova2025} (in brackets). The last column is the relative chi-square values for  $\bar{s}_{p,sys} $, $\bar{s}_{p,2} $  to the value  $\chi _{\bar{s}_{p,1} }^{2} $. for PR MSR expression given by Eq.(\ref{eq:9}).
}
\label{tab:tab2}
\begin{ruledtabular}
\begin{tabular}{ccc}
Expression  & Parameters  & $\bar{\chi }_{\bar{s}_{p,\alpha } }^{2} /\bar{\chi }_{\bar{s}_{p,1} }^{2} $ \\ \hline
$\bar{s}_{p,sys} $, (\ref{eq:12}) & $\delta ={\rm 0.37}, \gamma ={\rm 0.55}$($\delta ={\rm 0.37}, \gamma ={\rm 0.78}$) & 0.45(0.37) \\
$\bar{s}_{p,2} $, (\ref{eq:12}) & $\delta =0.435,  \gamma ={\rm 1.0}$($\delta =0.435,  \gamma ={\rm 1.0}$) & 0.85(1.45) \\
\end{tabular}
\end{ruledtabular}
\end{table}

As is seen from Table \ref{tab:tab2}, experimental data and microscopic calculations for the ratio $\bar{s}_{p} $ of fractions of PDR to GDR in E1 EWSR are statistically better described by systematics (\ref{eq:12}) than analytical  expressions PR MSR.

\begin{figure}[htb]
\centering
\setlength{\abovecaptionskip}{0pt}
\setlength{\belowcaptionskip}{8pt}
\includegraphics[scale=0.48]{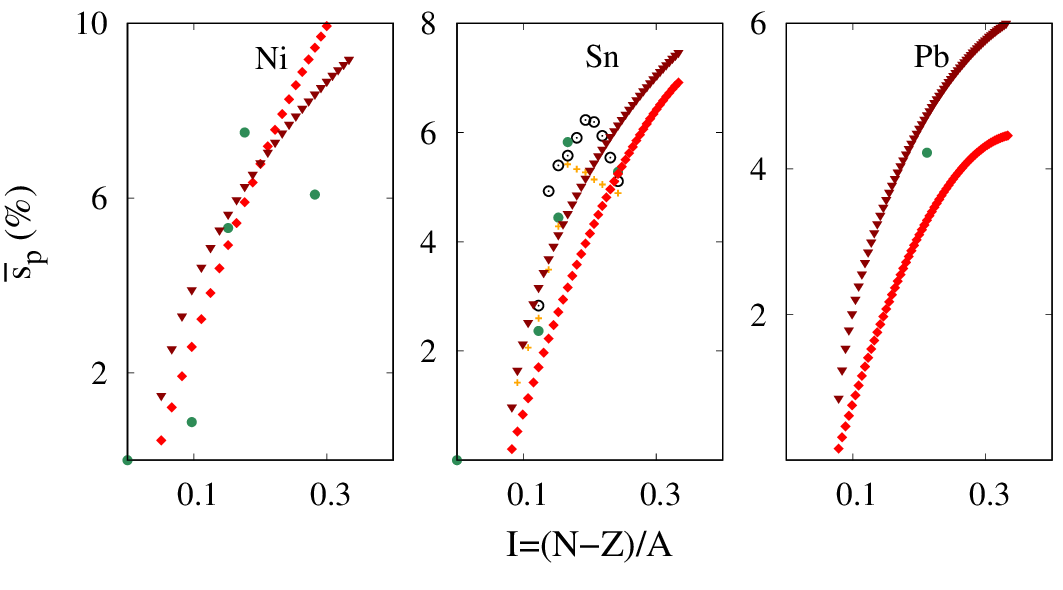}
\caption{
The values $\bar{s}_{p,1} $ (\ref{eq:9}) and $\bar{s}_{p,sys} $ (\ref{eq:12}) of the ratio of  the PDR to GDR fractions in E1 EWSR  for isotopes of Ni, Sn, Pb  as a function of the neutron-proton asymmetry parameter in comparison with microscopic data.
Denotations:  rhombus   $\color{red}\blacklozenge$ - calculation by expression (\ref{eq:9}) of PR MSF,
inverted triangle $\color{25}\blacktriangledown$ - systematics (\ref{eq:12}) for microscopic calculations ($\delta ={\rm 0.37}, \, \gamma ={\rm 0.55}$),
filled circle $\color{MyDarkTeal}\CIRCLE$  -  RRPA \cite{Vretenar2001},
circle $\odot$ - HB RQRPA \cite{Paar2003},
cross $\color{orange}+$ - MF RRPA \cite{Piekarewicz2006}.
}
\label{fig:fig5}
\end{figure}

Figure \ref{fig:fig5} demonstrates the dependence of  the ratio  $\bar{s}_{p} $ of the PDR to GDR sum rules for isotopes of Ni, Sn, Pb on the parameter of neutron-proton asymmetry $I=(N-Z)/A$.  The results of the calculations by PR MSR expression Eq.(\ref{eq:9})  and systematics (\ref{eq:12}) of microscopic calculations ($\delta ={\rm 0.37}, \, \gamma ={\rm 0.55}$) are presented in comparison with the microscopic results within RRPA \cite{Vretenar2001}, HB RQRPA \cite{Paar2003}, MF RRPA \cite{Piekarewicz2006}.

 Figure \ref{fig:fig5}  shows that the ratio of the PDR to GDR sum rules in accordance with the PR MSR expression (\ref{eq:9}) does not exceed $\approx$6-8\%
 in neutron-rich nuclei with $A\ge \approx 100$. These values are less in  $\approx$20-30\%
 than the values of microscopic calculations and systematics (\ref{eq:12}) of the microscopic calculations.  The values of the fractions of the PDR to GDR sum rules calculated by the PR MSR  expression  (\ref{eq:9}) and systematics   (\ref{eq:12}) of microscopic data are monotonically increasing with an excess of neutrons. In microscopic calculations such behavior of $\bar{s}_{p} $ is demonstrated at a small neutron excess.

\begin{figure}[htb]
\centering
\setlength{\abovecaptionskip}{0pt}
\setlength{\belowcaptionskip}{8pt}
\includegraphics[scale=0.48]{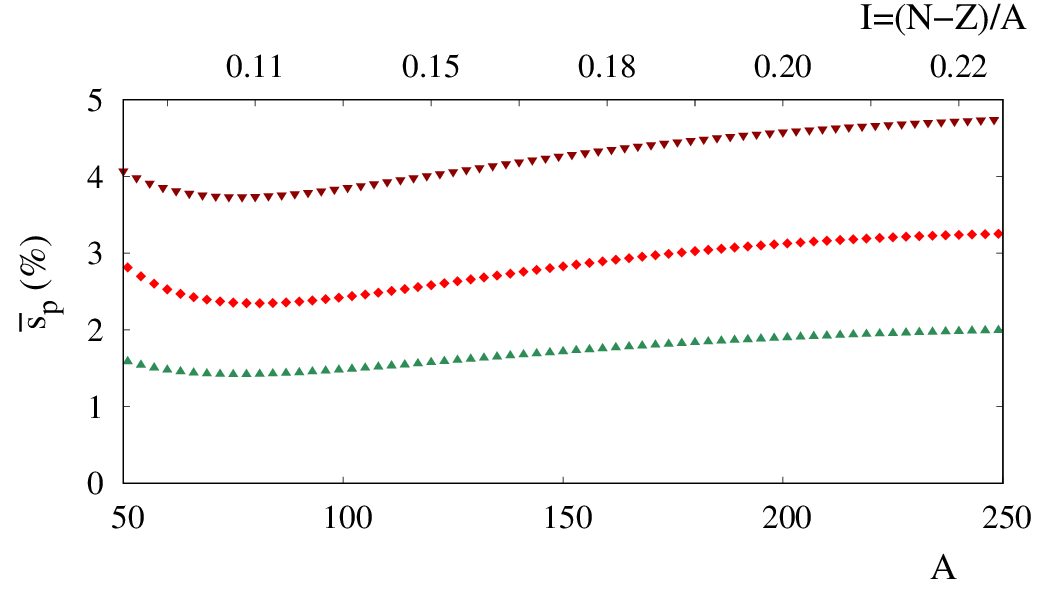}
\caption{
Ratio $\bar{s}_{p} $ of the PDR to GDR sum rules of E1 EWSR along the beta-stability line. 
Denotation: rhombus $\color{red}\blacklozenge$ - calculation by expression (\ref{eq:9}) of PR MSR,   
triangle $\color{MyDarkTeal}\blacktriangle$ - systematics (\ref{eq:12})  of the experimental data ($\delta ={\rm 0.37}$,  $ \gamma ={\rm 0.78}$), 
inverted triangle $\color{25}\blacktriangledown$ - systematics (\ref{eq:12}) of  microscopic calculations ($\delta ={\rm 0.37}$, $\gamma ={\rm 0.55}$).
}
\label{fig:fig6}
\end{figure}

 Figure \ref{fig:fig6} shows the ratio $\bar{s}_{p,sys} $ of systematics of the PDR to GDR sum rules of E1 EWSR along the beta-stability line as a function of the atomic mass number and parameter of neutron-proton asymmetry $I=(N-Z)/A$.  Ratio of the fractions is calculated according to PR MSF expression (\ref{eq:9}) and systematics (\ref{eq:12}) with the parameters from Table \ref{tab:tab2} of fitting the results of microscopic calculations and experimental data. The beta-stability line is determined by the Green formula \cite{Green1951,Green1952}: $I=0.4A/(200+A)$ .

 It can be seen that along the beta-stability line, the values of the ratio $\bar{s}_{p} $ of the PDR to GDR sum rules calculated within different methods do not exceed $\approx$5\%.
 The magnitude of $\bar{s}_{p} $  is placed between values for systematics for microscopic and experimental data. The values of the systematics of  microscopic calculations are placed higher than for  the systematics of experimental data.

\begin{figure}[htb]
\centering
\setlength{\abovecaptionskip}{0pt}
\setlength{\belowcaptionskip}{8pt}
\includegraphics[scale=0.48]{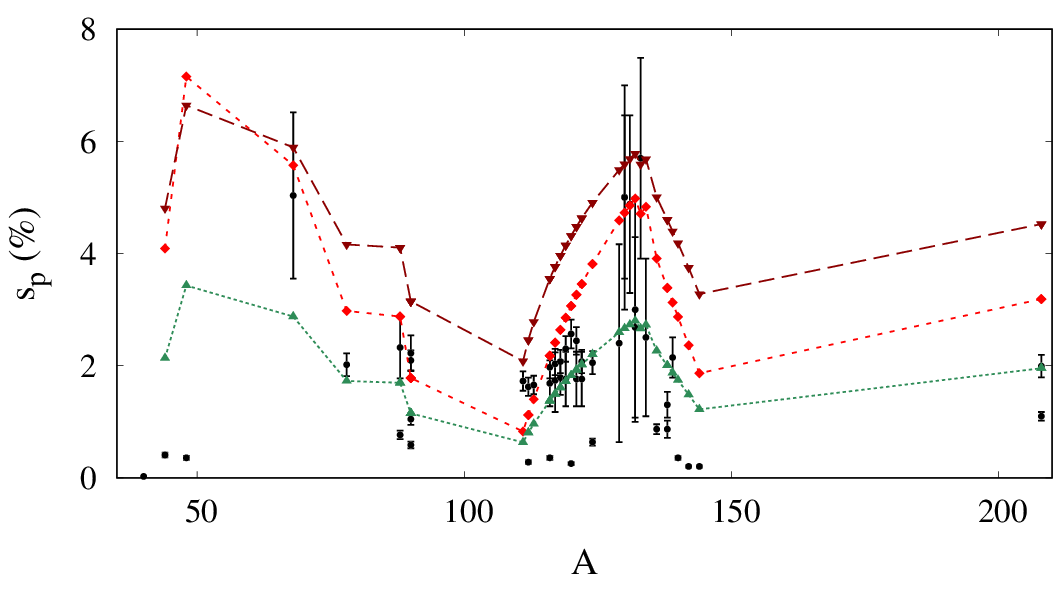}
\caption{
The comparison of the experimental data for the PDR fractions of E1 EWSR with calculations within different methods. 
Denotations: black points (with or without uncertainties) - experimental data, 
rhombus $\color{red}\blacklozenge$ - $s_{p,1} $ calculation by PR MSR, Eqs.(\ref{eq:11}), (\ref{eq:9}),
triangle $\color{MyDarkTeal}\blacktriangle$ - systematics (\ref{eq:13}), (\ref{eq:12}) with the parameters $\delta ={\rm 0.37}$, $\gamma ={\rm 0.78}$ by fitting experimental data,
inverted triangle $\color{25}\blacktriangledown$ - systematics (\ref{eq:13}), (\ref{eq:12}) with the parameters $\delta ={\rm 0.37}$, $\gamma ={\rm 0.55}$ by fitting microscopic calculations. 
The lines are drowned for  better visualization of the corresponding data.
}
\label{fig:fig7}
\end{figure}

 Figure \ref{fig:fig7} compares the experimental data of the PDR fraction $s_{p} $ of E1 EWSR with calculations within both  PR MSR,  Eqs. (\ref{eq:11}), (\ref{eq:9}),  and  systematics (\ref{eq:13}) with (\ref{eq:12})  of experimental and microscopic data. Experimental data are taken from Fig.26 of Ref.\cite{Savran2013} and Fig.2 of Ref.\cite{Markova2025}.  Data in Ref.\cite{Savran2013} were collected from  
 Refs.\cite{Govaert1998,Adrich2005,Savran2011,Hartmann2004,Klimkiewicz2007,Schwengner2007,Schwengner2008,Ozel2007,Volz2006,Makinaga2010,Schwengner2010,Schramm2012,Wieland2009,Poltoratska2012,Iwamoto2012,Toft2011}.

One can see that for the experimental data and the calculations within different approaches, the contributions of the PDR sum rule of E1 EWSR do not exceed $\approx$5\% 
in neutron-rich nuclei with $A\ge \approx 100$. The magnitudes of $s_{p} $  are placed between values for the systematics for microscopic and experimental data. 
The values of the systematics of microscopic calculations  are higher than for  the systematics of experimental data. 
It should be noted that there are ambiguities in the values of PDR sum rules for the experimental data and microscopic calculations in neutron-rich nuclei. 
The low-energy part of the excitation of the isoscalar giant dipole resonance is placed near isovector E1 PDR and the experimental data can be a mixture of E1  isovector and isoscalar excitations especially in ($\alpha, \alpha\prime$) experiment. 
The theoretical values of the PDR sum rule that were used for their systematics are also depent on the energy interval, over which they are determined. Here we considere that these effects are small from statistical point of view.

 Note, that in the PR INW approach, the value of PDR energy (\ref{eq:3}), (\ref{eq:4}) is a finite in the absence of the neutron skin like in the INW method for $N_{s} =0$ that is in contrast to the model SIS Eq.(\ref{eq:A26}). The fraction of the PDR to the E1 EWSR equals to zero in these situations, and there is no excitation of the PDR.

\section{Summary and conclusions }
\label{results}

The relationship for the PDR energy based on the macroscopic model of Isacker-Nagarajan-Warner \cite{Isacker1992} with the number of near-surface neutrons related to the thickness of the neutron skin is presented. 
The expression by Pethick- Ravenhall \cite{Pethick1996} is used for the calculation of the neutron skin (PR INW model). 
Such modification of the  macroscopic approach  by Isacker-Nagarajan-Warner enables to take into account microscopic observation  of direct relationship between skin thickness and low-energy  dipole response.

The neutron skin thickness is approximated by the linear function of the neutron excess parameter  $I=(N-Z)/A$  with coefficients depending on the atomic mass number $A$.  Within the framework of this macroscopic  PR INW model it is demonstrated that the values of the PDR energies for isotopes of Ni, Sn and Pb are consistent with microscopic calculations. In these situations, the  strength of effective proton-neutron interaction was taken to be equal to the value of the constant component $t_{0} $ = $\kappa _{np} =- 1828.23\,  {\rm MeV}\cdot \, {\rm fm}^{{\rm 3}}$ of the central interaction of the Skyrme force  MSk7 \cite{Goriely2001}. Simple expressions of systematics for the PDR energies in neutron-rich atomic nuclei are proposed and tested. The results agree rather well with those for PR INW, experimental data and microscopic calculations.

The PDR fraction of the E1 energy-weighted sum rule (integrated photo-absorption cross-section for E1 gamma quanta) is considered for the medium-heavy and heavy atomic nuclei. The systematics of the PDR fractions of E1 EWSR is proposed on the basis of the molecular energy-weighted E1 sum rule with a number of near-surface neutrons considered as a function of the thickness of the neutron skin (PR MSR model).  The comparison of the experimental data with the calculations within the framework of PR MSR, systematics and microscopic results are given. It demonstrates that contributions of the PDR sum rule to E1 EWSR do not exceed $\approx$ 5\%
in medium-heavy and heavy  neutron-rich nuclei.  The presented systematics of the  PDR characteristics can be used for the  estimation of the magnitudes of the corresponding quantities  in the absence of experimental data and microscopic calculations.

It should be noted that that microscopic results and experimental data for the PDR energies in the medium-heavy and heavy nuclei with neutron excess can be described by the INW PR approach as a whole if the absolute value of the strength of the neutron-proton interaction ($|\kappa _{np} |=1828.23\, \, {\rm MeV}\cdot \, {\rm fm}^{{\rm 3}} $) is three times as large  as that  ($|\kappa _{np} |=555.0\, \, {\rm MeV}\cdot \, {\rm fm}^{{\rm 3}} $) obtained by Isacker-Nagarajan-Warner by the volume integral of the nucleon-nucleon interaction. Therefore, while the macroscopic INW PR model can describe the main features of the PDR, above mentioned discrepancy of the strength values doesn't not provide reason enough for the conclusion that PDR is pure collective state. This question can be resolved by the microscopic methods in our opinion.

\begin{acknowledgments}
The authors are grateful to V. Yu. Denisov for important and constructive discussions on definitions and descriptions of macroscopic geometric characteristics of atomic nuclei.

V.A.P and O.M.G also thank the National Research Foundation of Ukraine  for support in part by grant No. NRFU 2023.05/0024 "Solving modern problems of chemistry, biomedicine, physics and materials science using the center of high-performance computing and machine learning".
\end{acknowledgments}

\appendix*

\section{Macroscopic models for the PDR energy and their parameters}

 The relationship for the square of the PDR energy by Isacker-Nagarajan-Warner approach \cite{Isacker1992} is based on Goldhaber-Teller mechanism of the GDR description \cite{Goldhaber1948}.  It can be obtained  from Eq. (\ref{eq:9}) in Ref.\cite{Isacker1992} and the equation of the frequency of the neutron surface oscillations relative to the nuclear core, and  can be presented in the form

\begin{multline}
\displaystyle E_{p}^{2} =\frac{4\pi |\kappa _{np} |}{3} \frac{\hbar ^{2} \rho _{n0} \rho _{p0} }{\mu _{p} } \Delta F(R_{n} ,R_{p} ),
 \\
\displaystyle  \Delta F=F(R_{n}, R_{p})-F(R_{p} ,R_{p} ).
\label{eq:A1}
\end{multline}

Here, $|\kappa _{np} |$  is the absolute value of the strength of effective neutron- proton interaction; $\mu _{p} $$=mN_{s} A_{c} /A$ is the reduced mass of the oscillating neutron-proton subsystems in the PDR; $N_{s} $ - a number of neutrons in the surface layer; $A_{c} =A-N_{s} $ a mass number of nuclear core; $\rho _{n0} =N/V$, $\rho _{p0} =Z/V$  are the central values of the density of protons and neutrons in the nucleus with volume $V$ =$(4\pi /3)R_{0}^{3} $ with the sharp nuclear radius $R_{0} =r_{0} A^{1/3} $ for the density of distribution of nucleons with a sharp edge. One can get the following general expression for the PDR energy

\begin{multline}
E_{p} =\left[\frac{\hbar ^{2} }{m} \frac{|\kappa _{np} |}{A} \frac{3}{4\pi r_{0}^{6} } \right]^{1/2}  \left[\frac{Z}{A_{c} } \frac{N}{N_{s} } \right]^{1/2} \times
\\
\times\left[\Delta F(R_{n} ,R_{p} )\right]^{1/2} .
\label{eq:A2}
\end{multline}

The functions $F(R_{n} ,R_{p} )$, $F(R_{p} ,R_{p} )$  describe displacement between  the distributions of neutrons and protons. They are defined by Eqs. {\ref{eq:6}) in Ref.\cite{Isacker1992} which can be presented in the form

\begin{multline}
F(R_{n}, R_{p} )=\frac{1}{4a} {\rm csch}^{2} (\frac{y}{2a} ) \times
\\
\times
\left[\coth (\frac{y}{2a} )\frac{c_{n} -c_{p} }{a} -(c'_{n} +c'_{p} )\right].
\label{eq:A3}
\end{multline}

Here, $c_{i} =(R_{i}^{3} +\pi ^{2} a^{2} R_{i} )/3$, $c'_{i} \equiv c'(R_{i} ,a)=\partial c_{i} /\partial R_{i} $=$R_{i}^{2} +\pi ^{2} a^{2} /3$; and $R_{n} ,\, \, R_{p} $-  radii of neutron and proton density distributions in the form of two parameter Fermi functions (2pF distribution),

\begin{equation}
\rho _{j} (r)=\rho _{0,j} \cdot \left[1+\exp (\frac{r-R_{j} }{a_{j} } )\right]^{-1},
\label{eq:A4}
\end{equation}

\noindent on half of the magnitude;  the diffusenesses of the distribution of protons and neutrons are considered the same $a_{n} =a_{p} =a$. The parameter $y=R_{n} -R_{p} $ is the neutron skin thickness for 2pF nuclear distributions (2pF neutron skin thickness). For $y=0$,

\begin{equation}
F(R_{p}, R_{p} )=\frac{3R_{p}^{2} +(\pi ^{2} -6)a^{2} }{18a} .
\label{eq:A5}
\end{equation}

The functions ${\rm csch}(x)$, $\coth (x)$ in (\ref{eq:A3}) are the hyperbolic cosecant and hyperbolic cotangent.  These functions have singularity at the argument going to zero. We transform Eq.(\ref{eq:A3})  to the following  expression without the functions with such singularities

\begin{multline}
F(R_{n}, R_{p} )=\bar{F}+\frac{y^{2} }{72a} +\frac{1}{4a} H_{1} (\frac{y}{2a} )\times
\\
\times \left[H_{2} (\frac{y}{2a} )\frac{y}{a} (\bar{c}'+\frac{y^{2} }{12} )+\frac{y^{2} }{a} (\bar{F}+\frac{y^{2} }{72a} )\right]+
\\
+H_{2} (\frac{y}{2a} )\frac{1}{y} (\bar{c}'+\frac{y^{2} }{12} ),
\label{eq:A6}
\end{multline}

\noindent where $\bar{F}\equiv F(R_{m} ,R_{m} )$, $\bar{c}'=c'(R_{m} ,a)$  with the mean radius $R_{m} =(R_{n} +R_{p} )/2$. The functions $H_{1} (x)$ and $H_{2} (x)$  with $x=y/(2a)$ are equal to $H_{1} (x)=$ ${\rm csch}^{2} (x)-\frac{1}{x^{2} } =-\frac{1}{3} +\frac{x^{2} }{15} +O(x^{4} )$, $H_{2} (x)=\coth ^{2} (x)-\frac{1}{x^{2} } -\frac{x}{3} =-\frac{x^{3} }{45} +O(x^{5} )$, and they have no singularity at $x=0$.

One gets formulas in the second order of $y$: for functions  $F$

\begin{multline}
F(R_{p} ,R_{p} )=\bar{F}-\frac{y\cdot R_{m} }{6a} +\frac{y^{2} }{24a},
\\
\bar{F}=\frac{3R_{m}^{2} +(\pi ^{2} -6)a^{2} }{18a} \cong \frac{R_{m}^{2} }{6a},
\label{eq:A7}
\end{multline}

\begin{multline}
F(R_{n} ,R_{p} )=\bar{F}\cdot \left[1-\frac{y^{2} }{10a^{2} } +\frac{y^{2} }{120a\bar{F}} \right]\cong
\\
\cong\frac{R_{m}^{2} }{6a} \cdot \left[1-\frac{y^{2} }{10a^{2} } \right].
\label{eq:A8}
\end{multline}

As a result, we find the following expression  for the PDR energy (\ref{eq:A2})  in the second order of $y$

\begin{multline}
E_{p} =\left[\frac{\hbar ^{2} }{m} \frac{|\kappa _{np} |}{A} \frac{3}{4\pi r_{0}^{6} } \right]^{1/2} \left[\frac{Z}{A_{c} } \right]^{1/2} \times
\\
\times
\left[\frac{R_{m} }{6a} \frac{yN}{N_{s} } \left(1-\frac{y\cdot }{30a^{2} } \frac{3R_{m}^{2} +\pi ^{2} a^{2} }{R_{m} } \right)\right]^{1/2}.
\label{eq:A9}
\end{multline}

We calculate a value of 2pF neutron skin thickness \textit{y} using the values of a root mean-square (rms) thickness (\cite{Elton1958,Elton1961,Hasse1988,Warda2010}), i.e., the difference $\Delta r_{np} =R_{rms,n} -R_{rms,p} $ of roots of the mean-square   radii $R_{rms,j} $ =$\sqrt{<r^{2} >_{j} } $ (rms radii) for neutrons ($j=n$) and protons ($j=p$) distributions.

For the medium-heavy and heavy atomic nuclei, the relationship $a/R_{j} <<1$ is fulfilled, therefore we use below leptodermic expansion. One can  get the following expressions for the 2pF neutron skin thickness $y$ and the 2pF mean radius $R_{m} $ as the functions of the corresponding mean-square nuclear characteristics:

\begin{multline}
y=R_{n} - R_{p} \cong \sqrt{\frac{5}{3} } \cdot \left[1+\frac{7\pi ^{2} }{10} \left(\frac{a}{R_{rms,p} } \right)^{2} \right]\Delta r_{np} \cong
\\
\cong \sqrt{\frac{5}{3} } \cdot \Delta r_{np},
\label{eq:A10}
\end{multline}

\begin{multline}
R_{m} =\frac{R_{n} +R_{p} }{2} \cong \sqrt{\frac{5}{3} } \cdot \left[R_{rms,p} +\frac{\Delta r_{np} }{2} \right].
\label{eq:A11}
\end{multline}

The rms radius of the proton distribution $R_{rms,p} $ is calculated  according to Refs.\cite{Angeli1999,Angeli2004,Angeli2013}

\begin{equation}
R_{rms,p} =\left[r_{1} +\frac{r_{2} }{A^{2/3} } +\frac{r_{3} }{A^{4/3} } \right]\cdot A^{1/3},
\label{eq:A12}
\end{equation}

\noindent where the values of the parameters were obtained from the fit of experimental data: (in fm) $r_{1} =0.9071$, $r_{2} =1.105$, $r_{3} =-0.548$. These values  are rather close to those used in modified drop model \cite{Myers1983} (in fm):$r_{1} =0.891$ ,$r_{2} =1.394$,$r_{3} =-0.930$ . We  have used the value of $a=a_{p} \cong 0.57$~fm \cite{Elton1961} for the parameter of diffuseness.

In accordance with theoretical studies and fitting experimental data  (\cite{Trzcinska2001,Jastrzebski2004,Swiatecki2005,Friedman2007,Warda2009,Lukyanov2020,Zhang2021}), the rms thickness $\Delta r_{np} $ of the neutron skin  for  medium-heavy and heavy nuclei can be approximated by a linear function of the parameter of the neutron-proton asymmetry $I=(N-Z)/A$

\begin{equation}
\Delta r_{np} =\alpha +\beta \cdot I.
\label{eq:A13}
\end{equation}

In various studies, the values of the parameters $\alpha ,\, \, \beta $ are slightly different, but the average values are in rather close agreement. It was shown from fitting experimental data in Ref.\cite{Zhang2021}, that chi- square deviation $\chi ^{2} $ for the linear function (\ref{eq:A13}) has a smaller value if the parameters $\alpha ,\, \, \beta $ are different for various chains of isotopes. We have found the dependence of the parameters $\alpha ,\, \, \beta $ on the atomic mass number using the relationships for neutron and proton rms radii determined by the microscopic approaches  from Refs.\cite{Warda1998,Seif2015}. In accordance with these studies, we have obtained overall dependences of the parameters of $\alpha ,\, \, \beta $ on the  atomic mass number for the first order of $I$  in the following form


\begin{equation}
\begin{array}{c}
\displaystyle \alpha \equiv \alpha _{W} = A^{1/3} (a_{1} +a_{2} /A),
\\
\displaystyle  \beta \equiv \beta _{W} =b_{1} A^{1/3},
\end{array}
\label{eq:A14}
\end{equation}

\begin{equation}
\begin{array}{c}
\displaystyle \alpha \equiv \alpha _{S}  =A^{1/3} (\bar{a}_{1} +\bar{a}_{2} A^{2/3} +\bar{a}_{3} A^{-1/3} ),
\\
\displaystyle \beta \equiv \beta _{S} =\bar{b}_{1} A^{1/3}  + \bar{b}_{2} A,
\end{array}
\label{eq:A15}
\end{equation}

\noindent where the factors $a_{j} ,b_{j} $ and $\bar{a}_{j}, \bar{b}_{j} $  are constant.   These factors are obtained from fitting of the least squares method of the experimental data to Eqs. (\ref{eq:A10})-(\ref{eq:A13}). The experimental data are taken  from  mixed database, which includes the evaluated data for even-even nuclei from Ref.\cite{Zhang2021} and data for other nuclei from Refs.\cite{Trzcinska2001,Jastrzebski2004}.  The following results are found

\begin{multline}
\alpha =-0.007\pm 0.012,  \ \
\beta =0.85\pm 0.08 ,
\\
\chi ^{2} =1.85 ,
\label{eq:A16a}
\end{multline}

\begin{multline}
a_{1} =-0.039\pm 0.006,
a_{2} =1.8\pm 0.2,
b_{1} =0.29\pm 0.03,
\\
 \chi ^{2} =1.17,
\label{eq:A16b}
\end{multline}

\begin{multline}
\bar{a}_{1}  =-0.22\pm 0.07,
\bar{a}_{2} =0.003\pm 0.002,
\bar{a}_{3} =0.62\pm 0.18,
\\
\bar{b}_{1}  =0.47\pm 0.09,
\bar{b}_{2} =-0.009\pm 0.005,
\chi ^{2} =0.96.
\label{eq:A16c}
\end{multline}

Here, $\chi ^{2} $  is the mean- square deviation of  the theoretical values (\ref{eq:A13})-(\ref{eq:A15}) of the rms thickness from the experimental data,

\[
\chi ^{2} =\frac{1}{N_{point} -N_{par} } \sum _{k=1}^{N_{point} }\left(\frac{(\Delta r_{np,k} )_{th} -(\Delta r_{np,k} )_{\exp } }{\sigma (\Delta r_{np,k} )_{\exp } } \right)^{2},
\]

\noindent where $N_{point} $ is the number of experimental points for the thickness of the neutron skin, $N_{par} $-  number of  parameters,  $\sigma (\Delta r_{np,k} )_{\exp } $-  standard deviation  of data point $k$.

We calculate the number of surface neutrons $N_{s}$ as a function of neutron thickness by using the relationship by Pethick- Ravenhall \cite{Pethick1996} which for the medium-heavy and heavy atomic nuclei can be presented in the following form

\begin{equation}
\frac{N_{s} }{N} =\frac{3y}{R_{0} } =\frac{\sqrt{15} \Delta r_{np} }{R_{0} },
\label{eq:A17}
\end{equation}

\noindent  which was used at  a later in Refs.\cite{Abrosimov2009,Kolomietz2007,Kolomietz2008}.  
The mean radius $R_{m} =(R_{n} +R_{p} )/2$ in the Eqs.(\ref{eq:A6})-(\ref{eq:A9})  is approximated by the expression $R_{m} =r_{m} \cdot A^{1/3} $  with a constant value of the parameter $r_{m} $. 
The exact value of the mean radius can be calculated by the formula $R_{m} =\left[R_{rms,p} +\Delta r_{np} /2\right]\cdot \sqrt{5/3} $ . 
We obtain the magnitude of the $r_{m} $ to fit $r_{m} \cdot A^{1/3} $ to exact value of the mean radius with Eqs.(\ref{eq:A13}), (\ref{eq:A16c}) for rms thickness $\Delta r_{np} $. The least squares method  is used for fitting with the minimization of the mean square deviation $\chi _{m}^{2} $:

\begin{multline}
\chi _{m}^{2} =\sum _{j}^{}w_{j}  \left(r_{m} g_{j} -R_{m,j} \right)^{2} , \ \
g_{j} =A_{j}^{1/3},
\\
R_{m,j} =R_{m} (A_{j} ).
\label{eq:A18}
\end{multline}

The weight factor  $w_{j} $ of the deviation of data point $j$ is taken as  $1/\sigma _{j}^{2} $ with $\sigma _{j} =0.1\cdot R_{m,j} $.  The minimum $\chi _{m}^{2} $  is realized at the following  value of the parameter $r_{m} $ with standard deviation $\sigma _{r_{m} } $:

\begin{multline}
r_{m} =\frac{\sum _{j}w_{j} g_{j} R_{m,j}  }{\sum _{j}w_{j} g_{j}^{2}  } ,\, \, \, \, \, \, \, \, \,
\sigma _{{r}_{m} }^{2} =\frac{\sum _{j} \left[w_{j} g_{j} \right]^{2} \sigma _{j}^{2} }{\left[\sum _{j}w_{j} g_{j}^{2}  \right]^{2} } .
\label{eq:A19}
\end{multline}

The calculations were performed for atomic mass numbers $A_{j} $  along the beta-stability line with the proton numbers $20\le Z_{j} \le 93$.  The atomic  mass numbers are taken in accordance with Green formula $N-Z=0.4A^{2} /(A+200)$ \cite{Green1951,Green1952,Moller1995}, namely,

\begin{multline}
A_{j} =5\left[Z_{j} +\sqrt{Z_{j}^{2} +40Z_{j} +10000} -100\right]/3.
\label{eq:A20}
\end{multline}

As a result, the following  value for the parameter $r_{m} $ is obtained:

\begin{equation}
r_{m} \cong R_{m} /A^{1/3} ={\rm 1.25}\pm {\rm 0.02}~{\rm fm}.
\label{eq:A21}
\end{equation}

Note that, the dependence of the GDR energy on atomic mass number  for the method by Isacker - Nagarajan-Warner \cite{Isacker1992} at constant $r_{m} $ is proportional to $A^{-1/6} $, i.e., it corresponds that one for Goldhaber-Teller model \cite{Goldhaber1948,Denisov2025}, if the parameter $r_{m} =R_{m} /A^{1/3} $ is a constant. Indeed, the relationship for the GDR energy in the INW method has the following form

\begin{multline}
E_{g} =\left[\frac{1}{A} \frac{\hbar ^{2} }{m} |\kappa _{np} |\frac{3}{4\pi r_{0}^{6} } \right]^{1/2} [F(R_{n} ,R_{p} )\; ]^{1/2}.
\label{eq:A22}
\end{multline}

We obtain  the  GDR energy in the second order of thickness $y$

\begin{multline}
E_{g} =\left[\frac{\hbar ^{2} }{m} |\kappa _{np} |\frac{3}{4\pi r_{0}^{6} } \right]^{1/2} \left[\frac{F(R_{n} ,R_{p} )}{A} \right]^{1/2}
\cong
\\
\cong  E_{GT} \cdot \left[1-\frac{y^{2} }{20a^{2} } \right],
\label{eq:A23}
\end{multline}

\noindent with

\begin{multline}
E_{GT} =\sqrt{\frac{\hbar ^{2} }{m} \, \frac{|\kappa _{np} |R_{m}^{2} }{8\pi ar_{0}^{6} A^{2/3} } } \cdot A^{-1/6} \cong
\\
\cong
\sqrt{\frac{\hbar ^{2} }{m} \, \frac{|\kappa _{np} |r_{m}^{2} }{8\pi ar_{0}^{6} } } \cdot A^{-1/6} =K_{g} \cdot A^{-1/6}.
\label{eq:A24}
\end{multline}

For values $a=0.57$~fm, $r_{0} =1.15$~fm \cite{Myers1983}, $r_{m} =1.25$~fm (\ref{eq:A21}), $\hbar ^{2} /m \simeq \hbar ^{2} /m_{proton} $ $\simeq 41.5$~${\rm MeV}\cdot {\rm fm}^{2} $ and $|\kappa _{np} |= 1828.23$~${\rm MeV}\cdot {\rm fm}^{3}$, the factor  $K_{g} $ in (\ref{eq:A24}) equals to

\begin{equation}
K_{g} =1.40\sqrt{|\kappa _{np}|} =59.9 \ \   {\rm MeV} .
\label{eq:A25}
\end{equation}

The expression for the PDR energy was first proposed by Suzuki-Ikeda-Sato in Ref.\cite{Suzuki1990}. It can be presented in the following form (in MeV)

\begin{multline}
E_{p,SIS} =\frac{2.082}{A^{1/3} } \left[\frac{\hbar ^{2} 8a_{sym} }{m\,  r_{0}^{2} }
\frac{ZN}{A^{2} } \right]^{1/2} \left[\frac{N_{s} }{A-N_{s} } \,
\frac{Z}{N} \right]^{1/2} =
\\
=79\left[\frac{4ZN}{A^{2} } \right]^{1/2} \left[\frac{N_{s} }{A-N_{s} } \, \frac{Z}{N} \right]^{1/2} A^{-1/3},
\label{eq:A26}
\end{multline}

\noindent where the values are used: $\hbar ^{2} /m\simeq \hbar ^{2} /m_{proton} $ $\simeq 41.5$ ~ ${\rm MeV}\cdot {\rm fm}^{2}$, $r_{0} =1.15$~fm, $a_{sym} =23$~MeV.

\bibliography{main}

\end{document}